\documentclass[%
aip,
amsmath,amssymb,
reprint,%
]{revtex4-1}
\usepackage{graphicx}
\usepackage{dcolumn}
\usepackage{bm}
\usepackage[utf8]{inputenc}
\usepackage[T1]{fontenc}
\usepackage{mathptmx}
\begin{document}
	
\title{Evaluation of the spectrum of a quantum system using machine learning based on incomplete information about the wavefunctions.}

\author{Gennadiy Burlak}
\email{gburlak@uaem.mx.}
\affiliation
	{CIICAp, Universidad Aut\'{o}noma del Estado de Morelos, Av. Universidad 1001,
		Cuernavaca, Morelos 62210, M\'{e}xico.}

\begin{abstract}
We propose an effective approach to rapid estimation of the energy spectrum of
quantum systems with the use of machine learning (ML) algorithm. In the ML
approach (back propagation), the wavefunction data known from experiments is
interpreted as the attributes class (input data), while the spectrum of
quantum numbers establishes the label class (output data). To evaluate this
approach, we employ two exactly solvable models with the random modulated wavefunction
amplitude. The random factor allows modeling the incompleteness of information
about the state of quantum system. The trial wave functions fed into the
neural network, with the goal of making prediction about the spectrum of
quantum numbers. We found that in such configuration, the training process
occurs with rapid convergence if the number of analyzed quantum states is not
too large. The two qubits entanglement is studied as well. 
The accuracy of the test prediction (after training) reached 98
percent. Considered ML approach opens up important perspectives to plane the quantum 
measurements and optimal monitoring of complex quantum objects.
\end{abstract}
\maketitle

\textit{Introduction}. The applications of intelligent machines in various context of scientific research recently become an area of active investigations
\cite{PNAS:2018,Linglong:2018,Lunden:2011,JAP-Urban:2019,JAP-Damir:2018,Laura:2018,Juan:2017,Salvail:2013,PNAS:2018,Vogel:1989,Smithey:1993,Breitenbach:1997,White:1999,Hofheinz:2009,Steinbrecher:2019}.
 One of the important perspective directions of quantum physics is the measurement of wave functions and the energy spectrum of quantum objects. At present, the wavefunction is determined using the tomographic methods \cite{Vogel:1989}-\cite{Hofheinz:2009}, which evaluate the wavefunction that is most compatible with a diverse set of measurements. The indirectness of these methods compounds the problem of direct determining the wavefunction. To overcome this problem, it was shown \cite{Lunden:2011} that the photon wavefunction can be measured directly by sequentially measuring two additional system variables. As result, the components of the wavefunction appear directly on the measuring apparatus. The alternative approach can be used to determine the polarization quantum state \cite{Salvail:2013}. In contrast to various works motivated by physically oriented approaches to artificial intelligence, there are much fewer practical studies estimating the energy spectrum (spectral numbers) for a quantum system based on incomplete information about the wavefunction with the use of the machine learning (ML) approaches. The information on the quantum object structure normally is obtained as the result of measurements of the wavefunction in series of experiments. By the measurements we mean the definition values of wavefunction at a number of spatial points $x_{j}$, see Ref.\cite{Lunden:2011} Fig.2(a). According to the Heisenberg uncertainty principle, in quantum theory, an exact measurement of position $X$ violates the wavefunction of the particle and forces the subsequent measurement of momentum $P$ to become random. To include such a factor into consideration in the machine learning (ML), we model uncertainty as a random modulation of the measured amplitude of the wavefunction. Thus an incomplete wavefunction can be written as $\widetilde{\Psi}_{k}(x_{i}|n_{j})=\gamma\Psi(x_{i}|n_{j})$, where $k$ is the number of measurement, $n_{j}$ is a spectral number for $j-$state, and $\gamma$ is random valued variable. However, such ML way has not been realized hitherto. The purpose of this paper is to present an effective approach to solution of this problem. Herein, we create a neural network (NN) and investigate the use of the controlled ML method to generate predictions for the energy spectrum based on incomplete information on the wavefunction of a quantum system. To evaluate this approach, we employ two exactly solvable models with different structure of the energy spectrum and the different field polarization. 

\textit{Machine learning.}
To apply the supervised machine learning (ML) technique we first have to
construct the dataset $R$ to train the neural network (NN). Our
approach is based on the following strategy. (i) We calculate
the wavefunctions from the solution of the Schr\"{o}dinger equation (SE)
$H\Psi_{n}=E_{n}\Psi_{n}$, with $\Psi_{n}(x_{i})=\Psi_{ni}$\ in spatial points $x_{i}$
, $i=1..,I$, where $I$ is total number of points in which the value of the
wavefunction is measured. (Similar structure of $\Psi_{n}$ measurements was used in the experiment \cite{Lunden:2011}). (ii) For every spectral number $n_{j},j=1,2,...J$ ( $J$ is
the number of states in energy spectrum) we prepare the matrices
$R_{j}$ of sizes $I\times K$ constructed in following way. We multiply
the wavefunctions $\Psi_{ni}$ by random-valuated factor $\gamma$ (uniform
noise) that produces a set of perturbed wavefunctions $\Psi(x_{i}%
|n_{j})\rightarrow\widetilde{\Psi}_{k}(x_{i}|n_{j})\equiv\gamma\Psi
(x_{i}|n_{j})$ , where $k=1...K$, $k$ is number of the experiment and the variable $K$
(total number of experiments) is arbitrary parameter, normally we used $K\leq100$. Here the random factor
$\gamma$ simulates the uncertainness of the wavefunction amplitude in a real
experiment. In our study $\gamma$ is uniform quasi-random variable with large
period of repetition that is recalculated at every computing steps. In last
column $R_{j}$ it is inserted the spectral number $n_{j}$ corresponding to 
the energy spectrum of $\Psi_{n}$. (iii) Lastly, we
construct the large matrix $R=R_{1}\oplus$ $R_{2}\oplus$ ...
$R_{J}$ that serves as the input dataset for our NN. Such total matrix $R=R(\widetilde{\Psi}_{k}(x_{i}|n_{j}))$ has the following structure%
\begin{equation}
R=\left\{
\begin{array}
[c]{cccccc}%
\widetilde{\Psi}_{1}(x_{1}|n_{1}) & \widetilde{\Psi}_{2}(x_{2}|n_{1}) & ... &
... & \widetilde{\Psi}_{K}(x_{I}|n_{1}) & n_{1}\\
\widetilde{\Psi}_{K+1}(x_{1}|n_{1}) & \widetilde{\Psi}_{K+2}(x_{2}|n_{1}) &
... & ... & \widetilde{\Psi}_{2K}(x_{I}|n_{1}) & n_{1}\\
... & ... & ... & ... & ... & ...\\
\widetilde{\Psi}_{1}(x_{1}|n_{2}) & \widetilde{\Psi}_{2}(x_{2}|n_{2}) & ... &
... & \widetilde{\Psi}_{K}(x_{I}|n_{2}) & n_{2}\\
... & ... & ... & ... & ... & ...\\
\widetilde{\Psi}_{1}(x_{1}|n_{3}) & ... & ... & ... & ... & n_{3}%
\end{array}
\right\}  \label{Database}%
\end{equation}
It is notably, (a) that the randomly perturbed wavefunctions $\widetilde{\Psi
}_{k}(x_{i}|n_{j})$ in Eq.(\ref{Database}) already are not the exact solutions
of the Schr\"{o}dinger equation, however, they still relay to the spectral
spectral numbers $n_{j}$, and (b) since at training all the rows in the dataset
$R$ are randomly interchanged, the initial order of rows in $R$ is not
important. In the following Section we will use $R$ matrix as the input dataset, where
in accordance to ML terminology the values $\widetilde{\Psi}_{k}(x_{i}|n_{j})$
are defined as the attributes (features) class, while the spectrum $n_{j}$ is
defined as the label class (categorical values \cite{Schmidhuber:2015,Haykin:2009,Kelleher:2015}).   
\begin{figure}[ptb]%
\centering
\includegraphics[
width=0.45\textwidth
]%
{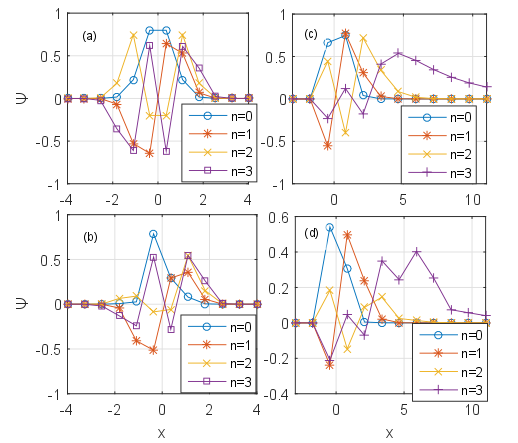}%
\caption{(Color on line.) (a) Spatial structure of the unperturbed
wavefunctions $\Psi_{n,j}$ (calculated for $12$ nodes) of the linear
oscillator for spectral numbers $n_{j}=0,1,2,3$ and $V_{0}=3$; (b) The
same as in (a) but for wavefunctions
$\protect\widetilde{\Psi}_{ni}$ randomly perturbed by uniform random noise with amplitude $A=1$;
(c) Spatial structure of the unperturbed wavefunctions $\Psi_{n,j}$
(calculated for $12$ nodes) of the no-symmetric Morse potential for
$V_{0}=7, \alpha=1$. In this case the condition $\sqrt{2V_{0}}/\alpha>n+\frac{1}{2}$
leads that only 4 discrete states $n_{j}=0,1,2,3$ are allowed. (d) The
same as in (c) but for randomly perturbed wavefunctions
$\protect\widetilde{\Psi}_{ni}$ by the uniform random noise.}%
\label{Fig1}%
\end{figure}
   
\textit{Neural network.}
We construct the NN and apply the ML technique to generate predictions of
spectral numbers $n_{j}$ that allow defining the energy\ spectrum of the
studied quantum system. For our ML studies we create a supervised NN by
standard technique \cite{Kelleher:2015} with the use of matrix $R$ from
Eq.(\ref{Database}). We use the back-propagation algorithm, which is one of
the common algorithms applied to train NN. In this paper, as a basic NN we used
the scheme from Ref.\cite{McCaffey:2015}, which, however, has been refined and considerably
expanded to reach our purpose: apply the numerical perturbed
wavefunctions $\widetilde{\Psi}_{nj}$ to predict the spectral numbers $n_{j}$. The values of the outputs $n_{j}$ are
determined by the input values $\widetilde{\Psi}_{nj}$, by the NN parameters: number of hidden
processing nodes that allow generating a set of weights and bias values. We created the
fully connected NN with $I_{L}$ inputs, $H_{L}$ hidden nodes and $O_{L}$
outputs that has $I_{L}H_{L}+H_{L}+H_{L}O_{L}+O_{L}$ weights and
biases. The next step is training of NN to calculate the values for the weights $W_{L}$ and biases $B_{L}$\ such that, for our
set of training data $R$ (with known input $\widetilde{\Psi}_{k}(x_{i}|n_{j}%
)$\ and output $n_{j}$\ values) the computed outputs of the NN has to
closely match the known outputs. In our simulation, the program splits the input dataset
$R$ randomly into two parts: training set
and test set. The training set (80\%) is used to create the neural network model, while
the test set (20\%) is used to estimate the accuracy of predictions. Besides of
the number of hidden nodes $H_{L}$ the following additional back-propagation
parameters are used in NN: the maximum number of training iterations $T_{\max
}$, the learning rate $L_{R}$ (that allows to controls the steepest descent
process), and the momentum rate $M_{R}$. The momentum rate helps to prevent
the training from the getting stuck with local, non-optimal weight values and
also prevents the oscillation where training never converges to stable values.
In realistic approach all the mentioned parameters are determined at
calculations by the trial and error. Also the program calculates the mean squared
error that allows to control the over-fitting at re-training\cite{Kelleher:2015}.

\textit{Exactly solvable models.}
To evaluate the applicability of the ML approach, we employ two exactly
solvable models with the random modulated wave functions. To do that we
consider the Schr\"{o}dinger equation (SE) solutions (to seek for
simplification we use here the atomic units allowing to set $m=1$ and $\hbar=1$)%
\begin{equation}
(1/2)d^{2}\Psi_{n}/dx^{2}+[E-V(x)]\Psi_{n}=0. \label{SchredEq}%
\end{equation}
The solutions of Eq.(\ref{SchredEq}) yield us the wavefunctions
$\Psi(x_{i}|n_{j})$ allowing to create the perturbed wavefunctions
$\widetilde{\Psi}_{k}(x_{i}|n_{j})=\gamma\Psi(x_{i}|n_{j})$ that are used in the input dataset $R$ in Eq.(\ref{Database}). Two well-known
models: linear oscillator and Morse potential\cite{LandauQM} are used here.

\textit{Linear oscillator.}
In this case in Eq.(\ref{SchredEq}) $V(x)=V_{0}x^{2}$ and the solution reads $\Psi_{n}=(\frac{\omega
}{\pi})^{1/4}\exp(-\frac{\omega}{2}x^{2})H_{n}(x\sqrt{\frac{\omega}{\pi}})$,
where $\omega=\sqrt{2V_{0}}$, $H_{n}(\xi)$ are the Hermit polynomials, and
$E_{n}=(n+\frac{1}{2})\omega$ is the energy for different spectral numbers
$n=0,...J<\infty$~ \cite{LandauQM}. Fig.\ref{Fig1} (a) displays the initial
wavefunctions with $V_{0}=3,$ $I=12$ (number of measurements points) and $J=4$
for states with $n=0,1,2,3$, while Fig.\ref{Fig1}(b) exhibits the
randomly perturbed functions $\widetilde{\Psi}_{\gamma}(x_{i}|n_{j})$, shown in Fig.\ref{Fig1}(a), but
modulated by random noise $\gamma$ \ (uniformly distributed random numbers in
$[0,...,1]$). 
\begin{figure}[ptb]%
\centering
\includegraphics[
width=0.45\textwidth
]
{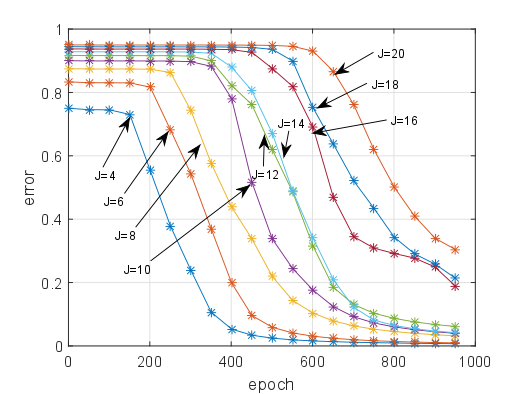}%
\caption{(Color on line) The value of the error training $\varepsilon_{T}$ as
function of the epoch numbers at training the linear oscillator dataset
(matrix $R$ of the perturbed wavefunctions $\protect\widetilde{\Psi}_{\gamma
}(x_{i}|n_{j})$) for different spectral numbers in the spectrum $J$ from $J=4$ (with $n=0,1,2,3$) to $J=20$ (with $n=0,1,2,...,19$).
We observe that the larger number of spectrum $J$ the larger number of epochs at training it is necessary to reach small errors.
}%
\label{Fig2}
\end{figure}
The random factor $\gamma$ is recalculated at every calculation
step, thus the spatial structure of $\widetilde{\Psi}_{k}(x_{i}|n_{j})$
changes significantly for every recalculating circle and it is different for
every rows (experiment $k$) in Eq.(\ref{Database}). The trial wavefunctions
$\widetilde{\Psi}_{k}(x_{i}|n_{j})$ fed into NN, with the goal of making
prediction about the spectrum of quantum numbers $n_{j}$. \ The parameters of
constructed NN may vary to improve the quality of predictions, initially we
used $I_{L}=I=12$ inputs (features), $J=O_{L}=4$ (outputs), and $H_{L}=5$
hidden layers, see Fig.\ref{Fig1} (a),(b). The appropriate choice
of the learning (back-propagation) parameters are assigned for the training
$T_{max}=1000$, $R_{L}=0.005$, $M_{L}=1$ , which are found to optimize the
efficiency of the NN model.   To monitor the dynamics of training we
calculated the mean squared error $\varepsilon_{T}=N^{-1}\sum_{i=1}^{N}%
(t_{i}-y_{i})^{2}$ that indicates a difference between the predicted value
$t_{i}$ and the true value $y_{i}$.
\begin{figure}[ptb]%
\centering
\includegraphics[
width=0.45\textwidth
]%
{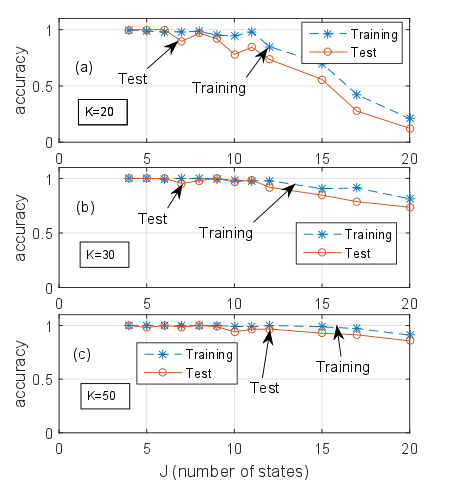}
\caption{(Color on line.) The accuracy of training $a_{tr}$ and accuracy of the test
$a_{ts}$ for the spectral numbers $n_{j}$ predictions  as function of the
maximal number of studied quantum states $J$ at different number of
experiments $K$, (a) $K=20$, (b) $K=30$, (c) $K=50$.
We observe that the accuracy $a_{ts}$ strongly depends on the quantity of experiments $K$. For small
$n_{j}<6$ the accuracy is high $a_{ts}\simeq95\%$\ already at $K=20$, but for large $n_{j}>10$
the accuracy $a_{ts}$ falls.
One can say the larger number of spectrum $J$ the larger number of experiments $K$ is necessary to reach the desired accuracy.}
\label{Fig3}%
\end{figure}
Fig. \ref{Fig2} shows the evolution of $\varepsilon_{T}$ as
function of the epoch numbers for the linear oscillator case. From Fig.
\ref{Fig2} one can see quite fast convergence of the training (for
not large spectral numbers $n$) that leads to small training errors
$\varepsilon_{T}\leq10^{-2}$ already for $500$ epochs. The convergence of the
training shown in Fig. \ref{Fig2} for not large $n<14$ brings the
test prediction accuracy for the linear oscillator model $a_{Ts}=$ $N_{c}/N$
about $\sim97\%$ and better (here $N_{c}$ is n\text{umber of correct
predictions, }$N$ is total n\text{umber of predictions)}. However the
prediction accuracy for large $n$ may be significantly improved at the use of
larger number of the epoch training $T_{max}>2000$.  
Fig. \ref{Fig3} exhibits the accuracy of training $a_{tr}$ and the accuracy of test
$a_{ts}$ for the spectral numbers $n_{j}$ predictions  as function of the   maximal number of studied quantum states $J$ at different number of experiments $K$, Fig. \ref{Fig3} (a) $K=20$, (b) $K=30$, (c) $K=50$.
We observe from Fig. \ref{Fig3} that the accuracy $a_{ts}$ strongly depends on the quantity of experiments $K$. For small $n_{j}<6$ the accuracy is high $a_{ts}\simeq95\%$\ already at $K=20$, but for large $n_{j}>10$ the accuracy $a_{ts}$ decreases. One can say the larger number of spectrum J the larger number of experiments is necessary to reach the desired accuracy. 
  
\textit{Morse potential.}
In this case in Eq.(\ref{SchredEq}) $V(x)=V_{0}(\exp(-2\alpha x)-2\exp(-\alpha x))$. In this potential case SE reads 
$\frac{d^{2}\psi}{dx^{2}}+2(E-A(\exp(-2\alpha x)-2\exp(-\alpha x))\psi=0$ with the solution $\psi=\exp(-\xi/2)\xi^{s}w(\xi)$ where $\xi=\frac{2\sqrt{2A}}{\alpha}\exp(-\alpha x)$ and $w$ can be expressed as a hypergeometric polynomial 
$w(\xi)=F(-n,2s+1,\xi)$. Here $s=\frac{2\sqrt{-2E}%
}{\alpha},n=\frac{\sqrt{2A}}{\alpha}-(s+\frac{1}{2})$. The latter allows the finite
energy spectrum $E_{n}=A[1-\frac{\alpha}{\sqrt{2A}}(n+\frac{1}{2})]^{2},n=0,1..<n_{max}$. 
Therefore it has to be $\frac{\sqrt{2V_{0}}}{\alpha}>n+\frac{1}{2}$, 
so there is not discrete spectrum if $\frac{\sqrt{2A}}{\alpha}<\frac{1}{2}$~ \cite{LandauQM}. 
Fig.\ref{Fig1}(c) exhibits the spatial structure of the unperturbed wavefunctions $\Psi_{n,j}$
(calculated for $12$ nodes) of the no-symmetric Morse potential for
$V_{0}=7, \alpha=1$. In this case the condition $\sqrt{2V_{0}}/\alpha>n+\frac{1}{2}$
leads that only 4 discrete states $n_{j}=0,1,2,3$ are allowed. Fig.\ref{Fig1}(d) displays
same as in  Fig.\ref{Fig1}(c), but for randomly perturbed wavefunctions by the uniform random noise
$\protect\widetilde{\Psi}_{ni}$ .      
\begin{figure}[ptb]%
\centering
\includegraphics[
width=0.45\textwidth
]%
{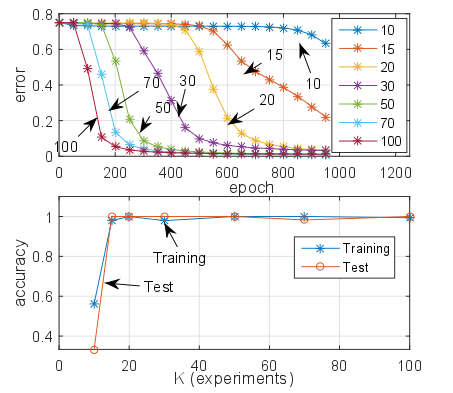}
\caption{(Color on line) (a) The training error $\varepsilon_{T}$ as function
of the epoch number at training the Morse dataset (matrix $R$ of the perturbed
wavefunctions) for parameters $V_{0}=7$, $a=1$ (in this case there is energy
spectrum with $n=0.1.2.3$). We observe that for ML parameters $R_{L}=0.005$,
$M_{L}=1$, $I_{L}=12$, $H_{L}=5$, $O_{L}=4$ the error $\varepsilon_{T}$
becomes small for the epoch iterations $T_{\max}\geq50$. (b) Training accuracy
$a_{Tr}$ and test accuracy $a_{Ts}$ as function of number of experiments $K$.}%
\label{Fig4}%
\end{figure}

Fig.\ref{Fig4} (a) shows the training error $\varepsilon_{T}$
as function of the epoch number at training of the Morse dataset (matrix $R$
of the perturbed wavefunctions $\widetilde{\Psi}_{\gamma}(x_{i}|n_{j})$) at
different $K$ for parameters $V_{0}=7$, $\alpha=1$ (in this case there is
only the limited energy spectrum with $n=0.1.2.3$). We observe that for ML
parameters $R_{L}=0.005$, $M_{L}=1$, $I_{L}=12$, $H_{L}=5$, $O_{L}=4$ the
error $\varepsilon_{T}$ becomes small for the epoch iterations at $T_{\max
}\geq50$. Fig.\ref{Fig4} (b) displays the training accuracy
$a_{Tr}$ and test accuracy $a_{Ts}$ as function of experiment number $K$. From
Fig.\ref{Fig4} (b) one can see that accuracy have
acceptable values $\geq95\%$ at quite large value of $K>40.$
\begin{figure}[ptb]%
	\centering
	\includegraphics[
	width=0.45\textwidth
	]%
	{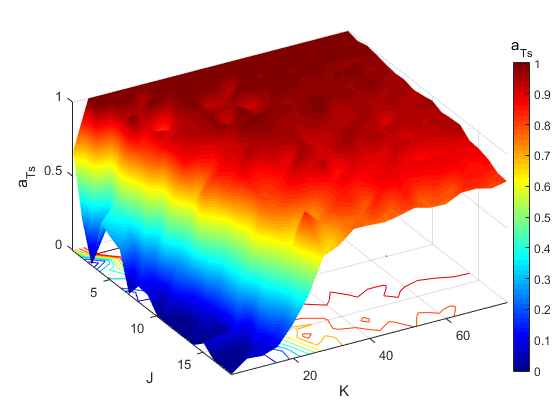}
	\caption{(Color on line) The test accuracy $a_{Ts}$ as function of the 
		spectrum photon numbers $J \in [2..20]$ and the numbers of experiment $K \in [5..100]$ at the spatial points $I=12$ and $1000$ epochs for polarized quantum states (photons) with two different wavelengths $\lambda_{1,2}$.}%
	\label{Fig5}%
\end{figure}

\textit{Advanced quantum model.} Now we consider an advanced quantum-mechanical model by adding the linear oscillator with characteristics of the orthogonal polarization state $p_1,p_2$ and the photon wavelength, that is a field state $\arrowvert{p_{1,2}},n,\lambda_{1,2}\rangle$, where $n$ is the numbers of photons radiated with two different wavelengths $\lambda_{1,2}$ (see Chap.14 in Ref. \cite{BasdevantQM:2005}). This allows us referring the experiments \cite{Lunden:2011,Salvail:2013}, where the wavefunction of photon was directly measured (see Figs.2,3 in \cite{Lunden:2011}). To study such a situation we add one more column (polarization+wavelength) to the features columns in the database $R$, see Eq.(\ref{Database}). Such a feature state $({p_{1,2}},\lambda_{1,2})$ is generated as random numbers in range $\in [1..4]$. 
Fig.\ref{Fig5} shows the test accuracy $a_{Ts}$ as function of the 
spectrum numbers $J \in [2..20]$ and the experiment numbers $K \in [5..100]$ at the fixed number of spatial points $I=12$ (where $\Psi$ was measured) and $T_{max}=1000$ epochs for polarized quantum states (photons)  $\arrowvert{p_{1,2}},n,\lambda_{1,2}\rangle$. One can see from Fig.\ref{Fig5} that NN for such the advanced model reaches the acceptable predictions with $a_{Ts}\simeq1$ only if the number of experiments is large enough $K>20$. 
\begin{figure}[ptb]%
	\centering
	\includegraphics[
	width=0.45\textwidth
	]%
	{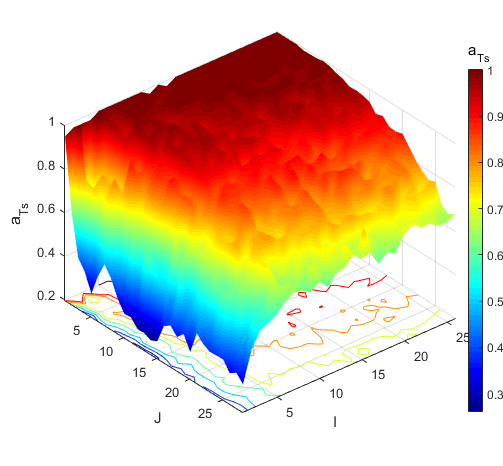}
	\caption{(Color on line) The test accuracy $a_{Ts}$ as function of the 
		 spectrum numbers $J \in [2..30]$ and the spatial points $I \in [4..30]$ at fixed number of experiment $K=100$ and $1000$ epochs for photons with wavelength $\lambda_{1,2}$.}%
	\label{Fig6}%
\end{figure}
Fig.\ref{Fig6} shows the test accuracy $a_{Ts}$ as function of the spectrum numbers $J \in [2..30]$ and the  spatial points $I \in [4..30]$ at fixed the experiment numbers  $K=100$ and $1000$ epochs for polarized  photons as in Fig.\ref{Fig5}. We observe from Fig.\ref{Fig6} that the acceptable predictions with $a_{Ts}\simeq1$ is reached only if a number spatial points $I$ is large enough $I>10$ and the spectrum photon numbers $J<10$.

\textit{Discussion.}
In this paper, we examined the use of NN to predict the energy spectrum (spectral numbers) and the photon entanglement 
from the measurements of the wave function of a quantum system. A database was built taking 
into account the number of measurements $J$ and the number of independent experiments on 
the training $K$. The effectiveness of such a network was demonstrated using two basic  
different solvable models of the Schr\"odinger equation. The first model (linear oscillator) 
has the unlimited energy spectrum, and in the second case (Morse potential) the number of 
energy levels depends on the amplitude of the potential. To model the uncertainty of wavefunctions, 
the random amplitude modulation was applied. The values of the wavefunction measurement were 
required only in a small number of spatial points $I = 12$ (Fig.\ref{Fig1}). Our calculations showed 
high efficiency of network training when the resulting learning error reached several percent 
for the number of epochs  less than $T_{max}=1000$ (Fig.\ref{Fig2}). The analysis of the spectral numbers 
prediction for both models showed that the training efficiency and the test accuracy depend 
significantly on the maximum quantity of spectral numbers $J$, and on the number of quantum 
experiments $K$. It was found that the large numbers of the spectrum $J$, the greater the number 
of epochs in machine learning is necessary to achieve an acceptable learning error. The larger $J$, 
the more experiments must be performed to achieve a desired accuracy of prediction shown in Figs \ref{Fig3},\ref{Fig4}. In the calculation of the train and test accuracy in Figs.\ref{Fig3},\ref{Fig4} the number of epoch kept fixed $T_{max}=1000$. 
In addition, the network predictions were studied for the \textit{Advanced quantum model} taking into account 
the difference in the polarization and field wavelength (polarization of photons), see Figs.\ref{Fig5},\ref{Fig6}. 
For such a system, it was also found that the accuracy of network predictions increases significantly 
with the increasing number of experiments, as well as the number of points, where the wavefunction was 
measured. Fig. 6 shows that when the length of the predicted spectrum is $J<15$, then value of the 
prediction accuracy practically does not increase with increasing of the measurement points number. 
Therefore, the number of the spatial measurement points of wavefunction $I=12$ used at the network 
analysis in Figs.\ref{Fig1}-\ref{Fig4} is optimal. As already noted, in view of the noticeable convergence of network 
training, the accuracy of predictions can be improved at increase of the number of training epochs. 
\begin{figure}[ptb]%
	\centering
	\includegraphics[
	width=0.55\textwidth
	]%
	{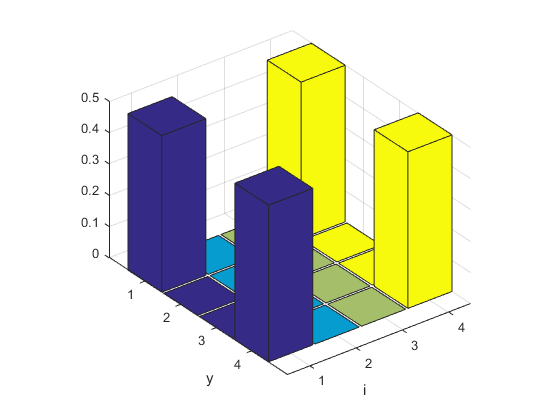}
	\caption{The density matrix $\rho_1^+ = \arrowvert\Phi^{+}\rangle\langle\Phi^{+}\arrowvert$ for the  polarized photon entangled states 
		(in the computing basis) obtained from our simulations for advanced model (see details in text).}%
	\label{Fig7}%
\end{figure}

The considered \textit{advanced model} has additional important features compared to the simple base models discussed above.
One of these features allow us to study a specific quantum entanglement for the mentioned qubits 
$\arrowvert{p_{i}},n,\lambda_{j}\rangle$ with $i=1,2; j=1,2$ by filtering them from the database $R$ created for the \textit{advanced} model. 
It is worth noting that, due to the orthogonality of quantum wavefunctions, only the states with the same spectral numbers $n$ contribute to the entangled qubit states.
In follows we will reformat the registered qubits to the Bell's states \cite{BasdevantQM:2005}
$(\arrowvert\Phi^{\pm}\rangle,\arrowvert\Psi^{\pm}\rangle)=\arrowvert{p_{i}},n,\lambda_{j}\rangle\pm\arrowvert{p_{i}},n,\lambda_{j}\rangle$
to prepare corresponding density matrices 
$\rho_1^\pm$ = ($\arrowvert\Phi^{\pm}\rangle\langle\Phi^{\pm}\arrowvert$, $\rho_2^\pm$ = ($\arrowvert\Psi^{\pm}\rangle\langle\Psi^{\pm}\arrowvert$. One of such
density matrices $\rho_1^+$ is shown (in the computing basis\cite{BasdevantQM:2005}) in Fig.\ref{Fig7}. 
In order to study the entanglement of such states we use a standard technique \cite{Wootters:1998} that allow us finding  a specific 
quantity: quantum concurrence $C$. The value $C=1$ indicates the complete two-qubit entanglement 
in the system. Our calculations (for details see \cite{Wootters:1998, Burlak:2009} and references therein) show that 
for considered states the value $C=1$, i.e. corresponding qubits pair are completely entangled.

\textit{Conclusion.} In our simulations, it is created and used the neural network with following typical parameters: $R_L = 0.005$, $M_L = 1$, 
input numbers $I_L = 12$, hidden nodes $H_L=5$. Our desktop with Core (TM), i3 processor, 8 GB RAM 
and 3.9 GHz processor, allowed us to achieve the test error $a_ {Ts}<2\%$ at the network 
training by the dataset with $12x2000$ entries for the CPU time less than $1min$. For large databases 
that were created and analyzed  for \textit{advanced} model (see 3D figures Figs.\ref{Fig5},~\ref{Fig6}) the run-time was about or less $10min$.  
Considered ML approach may open interesting perspectives at planning of the quantum measurements and the optimal 
monitoring of complex quantum objects.

\textit{Acknowledgment.}
This work was supported in part by CONACYT (M\'exico) under the grants No.
A1-S-9201 and No. A1-S-8793.

\end{document}